\documentclass[12pt]{iopart}

\pdfoutput=1

\usepackage{tikz}
\usepackage{graphicx}
\usepackage{amssymb}
\usepackage[mathscr]{eucal}

\newcommand{\affiliation}[1]{\address{#1}}

\begin{document}

\title{Perturbation Theory in k-Inflation Coupled to Matter}

\author{Philippe Brax\footnote{philippe.brax@cea.fr} and Emeline Cluzel\footnote{emeline.cluzel@cea.fr}}

\affiliation{Institut de Physique
Th\'eorique, CEA, IPhT, CNRS, URA 2306, F-91191Gif/Yvette Cedex, France  }

\date{today}

\begin{abstract}
We consider k-inflation models where the action is a non-linear function of both the inflaton and the inflaton kinetic term.
We focus on a scalar-tensor extension of k-inflation coupled to matter for which we derive a modified Mukhanov-Sasaki  equation for the curvature perturbation. Significant corrections to the power spectrum  appear when the coupling function changes abruptly along the inflationary trajectory.  This gives rise to a modification of  Starobinsky's  model of perturbation features. We analyse the way the power spectrum is altered in the infrared when such features are present.
\end{abstract}

\section{Introduction}
\label{sec:introduction}
DBI inflation \cite{Dvali:1998pa, Dvali:2001fw, Burgess:2001fx, Alexander:2001ks, Brodie:2003qv, KKLMMT, Silverstein:2003hf, Baumann:2009ds} is a well motivated alternative to slow roll inflation \cite{prems, KLS, Felder:1998vq}. Viewed as the low energy description of brane dynamics in an AdS throat, inflationary DBI branes are relatively  fast branes moving down along the throat \cite{KS, Kachru:2003aw}. In this context, the inflationary power spectrum and non Gaussianities have been investigated \cite{Garriga:1999vw, Easson:2009wc, Lorenz1, Lorenz2, Lorenz3, Alishahiha:2004eh, Bean:2007eh}.
An interesting extension of the DBI inflation scenario can be obtained when matter is present during inflation. In the brane context,
this occurs when trapped branes are present and stuck at fixed points of orbifold symmetries \cite{Green:2009ds}. In this case, the inflationary brane passes through trapped branes while particles living on the trapped branes are created. Such a creation may lead to a slowing down of the inflationary brane and therefore a modification of the background inflationary cosmology \cite{beauty,Brax:2009hd}.
In this paper, we focus on k-inflation models which generalise the DBI models with an action which becomes an arbitrary function of both the inflaton and its kinetic term. We consider the coupling of the inflaton to matter in k-inflation and its consequences on the perturbation theory.

A particularly remarkable situation arises when the coupling of the inflaton to matter is not constant but varies abruptly across a threshold value. The effective potential seen by the inflaton changes rapidly corresponding to a delta-like term in the effective mass of the inflaton.
Now, it is well known that such a delta function in the curvature perturbation equation leads to  features in the power spectrum \cite{Starobinsky:1992ts, Leach:2001zf}. We generalise this result to the case of k-inflation coupled to matter and find that the spectrum may be altered in the infrared.
Lately, many models of features in the CMB spectrum have been studied \cite{Adams:2001vc, Covi:2006ci, Hamann:2007pa, Joy:2007na, Hunt:2007dn, Joy:2008qd, Mortonson:2009qv, Dvorkin:2009ne, Hazra:2010ve}, either to explain   small features, or because they are  physically motivated in the context of standard inflation or non-canonical inflation \cite{Bean:2007hc, Bean:2008na, Barnaby:2010ke, Battefeld:2010rf}. Here we analyse  the power spectrum when the coupling function of the inflaton to matter changes abruptly during  k-inflation.

In a first part, we present the dynamics of k-inflation coupled to matter.
In section 3, we study the cosmological perturbation theory of these scalar-tensor theories. In section 4, we apply these results to the case of an abrupt change of the coupling to matter and eventually analyse the power spectrum of the curvature perturbation. We conclude in section 5.

\section{k-Inflation Dynamics}

\subsection{Brane scenario}

We will first recall some features of DBI inflation \cite{Silverstein:2003hf} in the presence of trapped branes \cite{beauty}. This will motivate the type of effective actions we will consider in the
rest of the paper.
The motion and the interaction of an inflationary DBI brane represented by the field $\phi$ with a trapped brane can be summarised using the effective potential
\begin{equation}
V(\phi,\chi)= m^2\phi^2 + g^2 \chi^2 (\phi-\phi_1)^2
\end{equation}
where $\phi_1$ is the position of the trapped brane and $g$ is a coupling constant. As the field $\phi$ gets close to $\phi_1$, particles of type $\chi$ living on the trapped brane are created.
The energy density of the created particles on the trapped brane can be written as \cite{beauty,Brax:2009hd}
\begin{equation}
\label{e2}
\rho_{\chi}= \frac{1}{(2\pi)^3} y(\xi)  \frac{a_{S}^{3}}{a^3} H^3 g \vert \phi -\phi_1\vert
\end{equation}
where $a_S$ is the scale factor at the end of the interaction zone. The
coupling function depends on $\xi$ where
\begin{equation}
\label{e3}
\xi=\frac{H^{2}}{g|\dot\phi|}
\end{equation}
is a constant for a quadratic potential
and reads
\begin{equation}
\label{e4}
y(\xi) \approx  {\xi}^{-3/2}
\end{equation}
when $\xi \ll 1$ as required to satisfy the Cosmic Microwave Background (CMB) COBE bound.
The created energy density leads to a modified  effective potential for the inflaton after the interaction region:
\begin{equation}
V_{\rm eff}\approx  m^2\phi^2 + \rho_{\chi}
\end{equation}
The time scale of the interaction region is
\begin{equation}
t_I\sim  (g\dot \phi)^{-1/2}
\end{equation}
implying that the creation of particles occurs on a much shorter time scale than one Hubble time.

The interaction of the trapped particles with the inflaton corresponds to  a $\chi$ field dependent mass
\begin{equation}
m_\chi(\phi)= g \vert \phi- \phi_1\vert
\end{equation}
As the size of the interaction region, when $\xi\ll 1$,  is given by
\begin{equation}
g\vert \phi_S -\phi_1\vert \approx \frac{g\vert \dot \phi\vert}{H},
\end{equation}
the mass of the created particles at the end of the interaction zone at $\phi_S$ becomes
\begin{equation}
m_\chi(\phi_S) \approx \frac{H}{\xi}\gg H
\end{equation}
Hence at the end of the interaction region, the trapped particles have a mass larger than the Hubble rate and therefore  behave like a fluid of Cold Dark Matter (CDM). This corroborates the fact that the density $\rho_\chi$ decays like $a^{-3}$ and is proportional to the mass of the $\chi$ particles in $\vert \phi -\phi_1\vert$.
Of course, the situation is drastically different in the interaction region where rescattering effects of the $\chi$ modes with the inflationary modes can lead to modifications of the
inflaton perturbations \cite{Barnaby:2010ke}.

In summary, the inflaton potential changes abruptly over a short time scale $Ht_I\ll 1$, going from
\begin{equation}
V(\phi)= m^2\phi^2
\end{equation}
before the passage through the trapped brane, as the $\chi$ field is very massive and stabilised at $<\chi>=0$, to an effective potential
\begin{equation}
\label{av}
V_{\rm eff}\approx  m^2\phi^2 + \frac{1}{(2\pi)^3} y(\xi)  \frac{a_{S}^{3}}{a^3} H^3 g \vert \phi -\phi_1\vert
\end{equation}
after brane crossing.

If we assume that there exists  a stack of N closely packed branes in the  interaction region, the minimum of the effective potential right after crossing the trapped brane is given in \cite{Brax:2009hd}
\begin{equation}
\frac{\phi_{\rm min}}{\phi_1} \approx  \frac{1}{16\pi^3} \frac{y(\xi) \epsilon\sqrt \lambda} {3g} N \frac{H^2}{M_p^2}
\end{equation}
This gives a criterion for the influence of the trapped brane on the motion of the inflationary brane. If $\phi_{\rm min}\ll \phi_1$, the trapped brane has no influence as the inflaton feels the $m^2\phi^2$ branch of the potential. On the other hand, if $H$ is large enough and $\phi_{\rm min}\gtrsim \phi_1$, the inflationary brane feels the steep potential due to the trapped brane. In this case, the motion of the inflationary brane is affected for a few e-foldings.
The slowing down is efficient if
\begin{equation}
\label{cond}
\frac{H^2}{M_p^2} \ge \frac{1}{N} 16\sqrt{3}\pi^3 g {\zeta^{1/2}}
\end{equation}
where $\zeta=\frac{g^2 \xi}{\sqrt{2} \pi}$ is the curvature perturbation on uniform-density hypersurfaces.
The COBE normalisation imposes \cite{Lorenz1}:
\begin{equation}
\label{upperH}
\frac{H}{M_p}\le 10^{-5}
\end{equation}
Unless we have at least $N\sim 10^9$ branes in the stack, the slowing effect of the stack is not drastic.

In the following, we will build an effective theory based on a k-inflation action coupled to non-relativistic matter with the same behaviour
of the inflaton potential as in the trapped brane case. We will study the perturbations in these k-inflaton models coupled to matter and analyse conditions for the existence of features which  generalise Starobinsky's results
when abrupt changes in the coupling to matter  are present.

\subsection{k-inflation coupled to matter}

The models we will consider are scalar-tensor theories where the inflaton field couples to matter.
In such  scalar-tensor theories, the action
is a sum of the k-inflation action for the inflaton, the matter term  and the Einstein-Hilbert action.
\begin{equation}
S=\int {\rm{d}}^4 x \sqrt {-g} \frac{m_{\rm{P}}^{2}}{2}R
+ \int{\rm{d}}^4 x \sqrt {-g} {\cal{P}}(\phi,X)
+\int{\rm{d}}^4 x
{\cal{L}}_{m}(\psi,{\tilde{g}}_{\mu\nu})
\end{equation}
in the Einstein frame, here $\psi$ is a matter field.
We have defined $X=\frac{1}{2}g_{\mu\nu}\partial^{\mu}\phi\partial^{\nu}\phi$
and the conformally $\phi$-dependent metric $\tilde{g}_{\mu\nu}=A^{2}(\phi)g_{\mu\nu}$.

Let us study  some general properties of the dynamics of k-inflation models coupled to matter.  First of all, the Einstein equations are not modified and read
\begin{equation}
R_{\mu\nu}-\frac{g_{\mu\nu}}{2}R= \kappa_4^2 (T_{\mu\nu}^\phi +T_{\mu\nu}^m)
\end{equation}
where $\kappa_4^2= m_{\rm P}^{-2}=8\pi G_N$. This a consequence of the non-modification of the Einstein-Hilbert term. The energy momentum tensors are defined by
\begin{equation}
T^{m,\phi}_{\mu\nu}= -\frac{2}{\sqrt{-g}} \frac{\delta S_{m,\phi}}{\delta g^{\mu\nu}}
\end{equation}
where we have identified
\begin{equation}
S_\phi= \int{\rm{d}}^4 x \sqrt {-g} {\cal{P}}(\phi,X)
\end{equation}
and
\begin{equation}
S_m=\int{\rm{d}}^4 x
{\cal{L}}_{m}(\psi,{\tilde{g}}_{\mu\nu})
\end{equation}
In terms of the non-linear function ${\cal P}(\phi,X)$, the inflaton energy momentum tensor is given by:
\begin{equation}
T_{\mu\nu}^\phi= g_{\mu\nu} {\cal P}- \frac{\partial {\cal P}}{\partial X} \partial_\mu \phi\partial_\nu \phi
\end{equation}
The dynamics of the inflaton are governed by the
Klein-Gordon equation
\begin{equation}
D_\mu \frac{\partial {\cal P}}{\partial_\mu \phi}- \frac{\partial {\cal P}}{\partial \phi}= \beta_\phi T^m
\end{equation}
where $T^m= g^{\mu\nu} T^m_{\mu\nu}$ is the trace of the matter energy momentum tensor and
\begin{equation}
\beta_\phi= \frac{\partial \ln A(\phi)}{\partial \phi}
\end{equation}
is the coupling constant of the inflaton to matter. The Klein-Gordon equation is also equivalent to
\begin{equation}
D_\mu \left(\partial^\mu \phi\frac{\partial {\cal P}}{\partial X}\right)- \frac{\partial {\cal P}}{\partial \phi}= \beta_\phi T^m
\end{equation}
Notice that due to the coupling to matter there is a new matter term in the Klein-Gordon equation.
The Bianchi identity implies that the total energy momentum is conserved
\begin{equation}
D^\mu (T^m_{\mu\nu}+ T^\phi_{\mu\nu})= 0
\end{equation}
But it also implies that matter is not conserved due to the energy exchange between matter and the inflaton
\begin{equation}
D^\mu T^m_{\mu\nu}= \beta_\phi T^m \partial_\nu \phi
\end{equation}
It is of particular interest to focus on
the case where  the matter fluid is pressureless
\begin{equation}
T^m_{\mu\nu}=\rho_E u_\mu u_\nu
\end{equation}
where the velocity field $u^\mu= \frac{dx^\mu}{d\tau}$ is normalised $u^2=-1$ and $\tau$ is the proper time along the
trajectories of the matter particles. The matter density $\rho_E$ is the Einstein frame density. It is not conserved as follows from
the (non-)conservation equation
\begin{equation}
\dot \rho_E +3 h \rho_E = \beta_\phi \rho_E \dot \phi
\end{equation}
where the time derivative along the trajectory is given by $\dot \rho_E= u^\mu D_\mu \rho_E$ and $3h = D^\mu u_{\mu}$ is the local expansion rate.
Defining
\begin{equation}
\rho_E= A(\phi)\rho_m
\end{equation}
we find that $\rho_m$ is conserved
\begin{equation}
\dot \rho_m +3h \rho_m=0
\end{equation}
In a cosmological context with the local Hubble rate $h=H$ being equal to the global one, we have that $\rho_m= \rho_0/a^3$.
Finally, the effect of the inflaton on the matter particles is to induce a scalar force as
\begin{equation}
\dot u_\mu= -\beta_\phi \dot \phi u_\mu -\beta_\phi \partial_\mu \phi
\end{equation}
This modifies the geodesics of the matter particles.

Coming back to the Klein-Gordon equation, we find that the dynamics are governed by an effective Lagrangian
\begin{equation}
\label{app1}
{\cal P}_{\rm eff}(\phi,X)={\cal P}(\phi,X)- \rho_m (A(\phi)-1).
\end{equation}
When the coupling is trivial $A(\phi)=1$, the effective Lagrangian is unchanged. It is only modified for non-trivial coupling functions.

\subsection{Abrupt changes in the coupling function}

When ${\cal {P}}(\phi,X)= P(\phi, X) - V(\phi)$, the effective potential seen by the inflaton is
\begin{equation}
\label{app2}
V_{\rm eff}(\phi)= V(\phi) + \rho_m (A(\phi)-1)
\end{equation}
where $\rho_m= \rho_0/a^3$ is the conserved energy density of the matter fluid.
This potential is similar to the ones used in chameleon models \cite{Brax:2004qh}.
We will  consider  the coupling function to be linear
\begin{equation}
A(\phi)= 1 + \frac{\vert \phi-\phi_1\vert}{\phi_1}Y_\delta (\phi_1 -\phi)
\end{equation}
where
the function $Y_\delta$ varies abruptly from 0 to 1  over a neighbourhood of the origin of size $\delta$ and is such that for $(\phi_1-\phi)>\delta$ we have
$Y_\delta (\phi_1- \phi) =1$.
Before the threshold value $\phi_1$, the coupling is identically $A(\phi)=1$ which implies that matter and the inflaton are effectively decoupled.
After the threshold crossing, the effective potential becomes
\begin{equation}
V_{\rm eff}(\phi)= V(\phi) + \rho_m \frac{\vert \phi-\phi_1\vert}{\phi_1}
\end{equation}
Now this is exactly the same change in the inflaton potential as the one in the trapped brane case when identifying
$P(\phi, X)=P_{\rm{DBI}}(\phi, X)$ and $\psi\equiv\chi$.
In the following, we will study
the consequences of such a change in the effective potential on the perturbation spectrum.

A simplified situation occurs when $\rho_m\ll V(\phi)$ implying that the matter density has no effect on the inflationary dynamics before the crossing of the threshold $\phi_1$. After the threshold, the effective potential has a matter dependent minimum where
\begin{equation}
\partial_\phi V\vert_{\phi_{\rm min}}= -\rho_m \partial_\phi A(\phi)\vert_{\phi_{\rm min}}
\end{equation}
Assuming that the potential $V(\phi)$ is a smooth function with a minimum at the origin, like $\frac{1}{2}m_0^2\phi^2$, the position of the minimum is determined by
\begin{equation}
\phi_{\rm min}= \frac{\rho_m}{m^2_0 \phi_1}
\end{equation}
where $m_0$ is the mass at the origin. As long as $\rho_m\ll m^2_0 \phi_1^2$, which amounts to neglecting the matter density compared to the inflaton energy density, we have $\phi_{\rm min}\ll \phi_1$ and the background dynamics of the inflaton are not influenced
by the coupling to matter.

\section{K-inflation Perturbation Theory}

\subsection{Cosmological dynamics}

The inflaton dynamics are governed by the energy density
\begin{equation}
\rho_\phi= -{\cal P} - \frac{\partial {\cal P}}{\partial X} \dot \phi^2
\end{equation}
and the pressure
\begin{equation}
p_\phi= {\cal P}
\end{equation}
In particular we find that
\begin{equation}
\frac{\partial {\cal P}}{\partial X}= \frac{\rho_\phi +p_\phi}{2X}
\end{equation}
The Klein-Gordon equation is also given by
\begin{equation}
\label{KGeff}
\frac{\partial {\cal P}_{\rm eff}}{\partial \phi} +\frac{\partial {\cal P}_{\rm eff}}{\partial X}(\ddot \phi + 3H \dot \phi) + \dot \phi \frac{d}{dt} (\frac{\partial {\cal P}_{\rm eff}}{\partial X})=0
\end{equation}
expressed as a function of the effective Lagrangian ${\cal P}_{\rm eff}$. In the following we will
 assume that the solutions to the Klein-Gordon equation lead to an inflationary behaviour and study cosmological perturbations of this inflationary background.

\subsection{Matter perturbations}
Due to the absence of anisotropic stress in the Einstein frame, we describe the metric perturbations using the Newton gauge, leading to the perturbed FLRW line element
\begin{equation}
\label{metric}
{\rm{d}}s^{2}=-(1+2\varphi_{N}){\rm{d}}t^{2}+a^{2}(t)(1-2\varphi_{N})\delta_{ij}{\rm{d}}x^{i}{\rm{d}}x^{j}
\end{equation}
where $\varphi_{N}$ is the Newtonian potential.
The velocity field can be expanded as
\begin{equation}
u_\mu=(-1-\varphi_N, v_i)
\end{equation}
Being interested in the scalar modes only we define
\begin{equation}
v_i=\partial_i \psi
\end{equation}
Calculating the local expansion rate to first order we find
\begin{equation}
3h=3H-3H\varphi_N-3\dot\varphi_N+\frac{1}{a^2}\partial_iv^i
\end{equation}
implying that the  Euler equation becomes
\begin{equation}
\dot \psi= -\varphi_N -\beta_\phi \dot \phi \psi -\beta_\phi \delta\phi
\end{equation}
Similarly the conservation equation reads
\begin{equation}
\dot \delta_m +\frac{\Delta \psi}{a^2}= 3 \dot \varphi_N
\end{equation}
These equations have to be complemented with the perturbed Einstein equations.

\subsection{Perturbed Einstein equations}

Following \cite{Bassett:2005xm} we define
\begin{equation}
\delta T^0_i\equiv \partial_i q
\end{equation}
with
\begin{equation}
q= - (\rho_\phi+ p_\phi) \frac{\delta \phi}{\dot \phi} +A\rho_m \psi
\end{equation}
This implies that
the 0i component of the perturbed Einstein equation leads to
\begin{equation}
\dot\varphi_{N}+H\varphi_{N}=4\pi G_N [(\rho_\phi+p_\phi)\frac{\delta\phi}{\dot\phi}-A\rho_m \psi]
\end{equation}
We also derive  the energy density perturbation
\begin{equation}
\delta T_{0}^{0}=-\delta\rho
\end{equation}
where $\rho= \rho_\phi + \rho_E$ is the total energy density.
Using
\begin{equation}
\delta \rho= \frac{\partial \rho}{\partial \phi} \delta\phi + \frac{\partial \rho}{\partial X} \delta X + A \delta\rho_m
\end{equation}
where $\delta\rho_m$ is the intrinsic matter perturbation and using
the conservation equation for the total energy density
\begin{equation}
\frac{d\rho}{dt}=-3H (\rho+p)= \frac{\partial \rho}{\partial \phi} \dot\phi + \frac{\partial \rho}{\partial X} \dot X + A\dot\rho_M
\end{equation}
we find that
\begin{equation}
\delta \rho= -3H(\rho_\phi+p_\phi)\frac{\delta\phi}{\dot\phi}+ \frac{\rho_\phi+p_\phi}{2c_s^2 X}\left(\delta X- \dot X \frac{\delta \phi}{\dot \phi}\right)+ A \delta\rho_m
\end{equation}
and finally
\begin{equation}
\delta \rho= -3H(\rho_\phi+p_\phi)\frac{\delta\phi}{\dot\phi}+ \frac{\rho_\phi+p_\phi}{c_s^2}\left(\frac{d}{dt}\frac{\delta\phi}{\dot\phi}-\phi_N\right)+ A \delta\rho_m
\end{equation}
We have defined the speed of sound as $c_s^2=\frac{\partial p/\partial X}{\partial\rho/\partial X}$ which only  depends on the inflaton.
With the metric (\ref{metric}), the perturbed Einstein tensor is
\begin{eqnarray}
-\frac{1}{2}\delta G_{0}^{0}&=&\frac{1}{a^2}\Delta\varphi_{N}-3H(\dot\varphi_{N}+H\varphi_{N})
\end{eqnarray}
which leads to the perturbed 00 Einstein equation
\begin{equation}
\label{zerozero}
\frac{\rm{d}}{\rm{d}t}\left(\frac{\delta\phi}{\dot\phi}\right)=\varphi_{N}+\frac{c_s^2}{4\pi G_N a^2(\rho_{\phi}+p_\phi)}\Delta\varphi_{N}+U_m
\end{equation}
where the source term from the matter perturbations is
\begin{equation}
U_m=3H\frac{c_s^2}{\rho_{\phi}+p_\phi}A\rho_m\psi -\frac{c_s^2}{\rho_\phi+p_\phi} A \delta\rho_m
\end{equation}
Together with the conservation and the Euler equations, these Einstein equations describe the system of perturbations.
They are valid for any k-inflation model coupled to matter.

\subsection{Curvature perturbation}

It is convenient to introduce  gauge invariant quantities and study their dynamical evolution.
The comoving curvature perturbation ${\cal R}_T$ is such  a gauge invariant quantity:
\begin{equation}
\label{defdeR}
{\cal R}_T=\varphi_{N}-\frac{H}{\rho +p} q
\end{equation}
This can be written as
\begin{equation}
{\cal R}_T= \varphi_N + H \frac{\rho_\phi+p_\phi}{\rho+p} \frac{\delta \phi}{\dot \phi} +{\cal R}_\psi
\end{equation}
where
\begin{equation}
{\cal R}_\psi= - \frac{A\rho_m H}{\rho+p} \left(\psi-\frac{\delta\phi}{\dot\phi}\right)
\end{equation}
We define
\begin{equation}
{\cal R}= {\cal R}_T + \frac{A\rho_m H}{\rho+p}\psi
\end{equation}
which coincides with the comoving curvature perturbation in the absence of matter
\begin{equation}
{\cal R}=\varphi_N + H\frac{\delta \phi}{\dot \phi}
\end{equation}
The effect of matter on ${\cal R}$ can be seen when analysing its time evolution
\begin{eqnarray}
\nonumber
\dot{\cal R}= H \frac{\Delta \varphi_N}{a^2} \frac{c_s^2}{4\pi G_N (\rho_\phi +p_\phi)} -4\pi G_N A\rho_m \frac{\delta \phi}{\dot \phi}
\\
+A\rho_m (-4\pi G_N + \frac{3H^2c_s^2}{\rho_\phi+p_\phi}) \psi  -\frac{c_s^2 H }{\rho_\phi+ p_\phi} A\delta \rho_m
\end{eqnarray}
It is  convenient to rewrite
\begin{equation}
\dot{\cal R}=C \frac{\delta \phi}{\dot \phi} + D R + T_\psi
\end{equation}
where we have identified
\begin{equation}
C= \frac{k^2H^2}{a^2} \frac{c^2_s}{4\pi G_N (\rho_\phi+p_\phi)} (1+ W),\ \ \ D=- \frac{k^2 H c_s^2}{a^2 4\pi G_N (\rho_\phi +p_\phi)}
\end{equation}
and
\begin{equation}
W=-\frac{k_c^2}{k^2}
\end{equation}
The characteristic momentum $k_c$ is given by
\begin{equation}
\label{kc}
k_c^2= A\rho_m \frac{a^2 (4\pi G_N)^2 (\rho_\phi +p_\phi)}{H^2 c_s^2}
\end{equation}
When matter is absent we have $W=0$.
We have also introduced
\begin{equation}
T_\psi= \left(-4\pi G_N + \frac{3H^2 c_s^2}{\rho_\phi+p_\phi}\right)A\rho_m\psi - \frac{Hc^2_s}{\rho_\phi+p_\phi} A \delta \rho_m
\end{equation}
This allows one to obtain  a second order differential equation for ${\cal R}$
\begin{equation}
\ddot{\cal R} + \left(H- \frac{\dot C}{C}\right)\dot {\cal R} +\left(\frac{k^2 {\tilde c}_s^2}{a^2} +4\pi G_N A\rho_m\right) {\cal R}={\Delta_\psi}
\end{equation}
or equivalently in conformal time defined by ${\rm{d}}t=a {\rm{d}}\eta$ and $'={\rm{d}}/{\rm{d}}\eta$.
\begin{equation}
{\cal R}'' - \frac{ C'}{C} {\cal R}' +({k^2 \tilde {c}_s^2}+ a^2 4\pi G_N A\rho_m){\cal R}= a^2{\Delta_\psi}
\end{equation}
where we have identified the effective speed of sound
\begin{equation}
\tilde {c_s^2}= c_s^2 \left(1- \frac{ W'}{{\cal H}(1+W)} \frac{2}{3(1+w_\phi)}\right)
\end{equation}
and we have used the inflaton equation of state $w_\phi=p_\phi/\rho_\phi$.
The source term reads
\begin{equation}
\Delta_\psi= \dot T_\psi + \left(H-\frac{k^2c_s^2}{4\pi G_N a^2(\rho_\phi+p_\phi)}\right) T_\psi  -\frac{\dot C}{C} T_\psi + C U_m
\end{equation}
Let us define
\begin{equation}
z_A = z \vert 1+ W\vert^{-1/2}
\end{equation}
where
\begin{equation}
z=  \frac{a(\rho_\phi+p_\phi)^{1/2}}{Hc_s}
\end{equation}
and the modified Mukhanov-Sasaki variable
\begin{equation}
v_A= z_A {\cal R},
\end{equation}
we then find that
\begin{equation}
v_A'' + \left(k^2 \tilde c_s^2 + a^2 4\pi G_N  A\rho_m -\frac{z_A''}{z_A}\right) v_A = a^2 z_A \Delta_ \psi
\end{equation}
When matter is absent, this reduces to the usual Mukhanov-Sasaki equation generalised to k-inflation by Garriga and Mukhanov \cite{Garriga:1999vw}.
In our case the Mukhanov-Sasaki variable 
 is k-dependent.

The full perturbation equations are very complex. Here we will simply emphasize some salient points which differ from the case with no matter.
First of all, the perturbation equations depend crucially on the scale $k_c$ which is time dependent. When $k\gg k_c$, the speed of sound is not altered $\tilde c_s=c_s$.
On larger scales when $k\ll k_c$, the speed of sound $\tilde c_s$ is largely modified by the presence of matter. Similarly, $z_A$ differs greatly from $z$ when $k\ll k_c$. Moreover, as matter perturbations enter as sources in the $v$-equation, we expect new modes which would affect the ${\cal R}_T$ power spectrum.

During an acceleration era such as the late time acceleration of the universe where $\rho_m$ and $\rho_\phi$ are of the same order, the equations are difficult to tackle analytically.
On the other hand, during primordial inflation when the number of inflationary efoldings is large, the influence of $\rho_m$ is limited to a few efoldings before being red-shifted away. In this case, modes of interest will always satisfy $k\gg k_c$. Moreover we can concentrate on the efoldings when $\rho_m\ll \rho_\phi$. Despite being negligible at the background level, the matter energy density can play a significant role when the matter coupling $A(\phi)$  varies abruptly along the inflationary trajectory. We will focus on this possibility in the following section.

\section{Features in the Power Spectrum}

\subsection{Starobinsky's model}

In this section we will recall the results due to Starobinsky \cite{Starobinsky:1992ts} in the case of a simple feature of the delta function type.
Let us consider a canonical model of inflation with
\begin{equation}
{\cal P}= -X -V(\phi)
\end{equation}
such that $V(\phi)$ is piece-wise linear
\begin{equation}
V(\phi) = V_0 + \left(A_- Y(\phi_1-\phi) + A_+ Y(\phi-\phi_1)\right) (\phi-\phi_1)
\end{equation}
where $Y$ is the Heaviside function. At the perturbation level  in a fixed gravitational background, the Klein-Gordon equation reads
\begin{equation}
\ddot\delta\phi + 3H \dot \delta \phi + (m^2_\phi- \frac{k^2}{a^2}) \delta \phi=0
\end{equation}
where $m^2_\phi= \frac{d^2V}{d\phi^2}$ is the mass of the inflaton. In Starobinsky's model, the mass essentially vanishes everywhere but for a $\delta$ function singularity
\begin{equation}
m^2_\phi= (A_+-A_-) \delta (\phi-\phi_1)
\end{equation}
Of course, when dealing with cosmological perturbations one cannot neglect the gravitational perturbations implying that the relevant equation is the Mukhanov-Sasaki equation which essentially reads
\begin{equation}
v'' +\left(k^2+  3{\cal H} \left(1-\frac{A_-}{A_+}\right) \delta (\eta-\eta_1) -\frac{a''}{a}\right) v=0
\end{equation}
Hence we see that $\delta$ function singularities do appear in cosmological perturbation equations. The solutions to this equation will be recalled in the following section.

Let us now consider a scalar-tensor model with canonical kinetic terms and the effective potential
\begin{equation}
V_{\rm eff}= V(\phi) + \rho_m (A(\phi) -1)
\end{equation}
where in the $\delta\to 0$ limit we take
\begin{equation}
A(\phi)= 1 + \frac{\vert \phi-\phi_1\vert}{\phi_1} Y(\phi_1 -\phi)
\end{equation}
The mass of the inflaton $m^2_\phi= \frac{d^2 V_{\rm eff}}{d\phi^2}$ then becomes
\begin{equation}
m^2_\phi= \frac{d^2 V}{d\phi^2} + \rho_m \frac{\delta (\phi-\phi_1)}{\phi_1}
\end{equation}
Even when the influence of $\rho_m$ on the background evolution is negligible, this $\delta$ function in the mass may play a role at the perturbation level.
Before analysing the effect of an abrupt change of $A(\phi)$ we will first recall the generic properties of the solution of the Mukhanov-Sasaki equation with
a $\delta$ function potential.

\subsection{The power spectrum}
In conformal time, we consider the perturbation equation for the Mukhanov-Sasaki variable $v$ in a quasi de  Sitter phase with a delta function feature at time $\eta_1$
\begin{equation}
v'' +\left(c_s^2 k^2 -\frac{z''}{z} +u\delta(\eta-\eta_1)\right)v=0
\end{equation}
In the following we will analyse the solutions when $c_s$ is constant. A slightly better approximation amounts to changing  adiabatically $c_s \to c_s(\eta)$ in the solutions as long as $c_s(\eta)$ is a slowly varying function. Such an approximation is acceptable at first order \cite{Lorenz3}.
To leading order in the slow roll parameters, the de Sitter term $\frac{z''}{z}\approx \frac{a''}{a}$ is a good approximation for the potential term in the
perturbation equation.

 It is convenient to define
$x= kc_s \eta$ then
\begin{equation}
\frac{d^2 v}{dx^2} +\left(1-\frac{2}{x^2} + \hat u \delta (x-x_1)\right)v=0
\end{equation}
whose solutions are
\begin{equation}
(\pm i +\frac{1}{x}) e^{\mp ix}
\end{equation}
with $\hat u= \frac{u}{kc_s}$. Notice that $\hat u$ is dimensionless.
Before the feature we have a Bunch-Davies vacuum with
\begin{equation}
v= {\cal C} \left(i+\frac{1}{x}\right) e^{-ix}
\end{equation}
where ${\cal C}\propto \frac{1}{\sqrt{2k}}$
and after the passage
\begin{equation}
v= \alpha \left(i+\frac{1}{x}\right) e^{-ix} +\beta \left(-i+\frac{1}{x}\right) e^{ix}
\end{equation}
with the junction condition
\begin{equation}
\left[\frac{dv}{dx}\right]_{x_1}= -\hat u v(x_1)
\end{equation}
The Bogoliubov coefficients are
\begin{equation}
\alpha= {\cal C}\left(1+\frac{\hat u}{2i}\left(1+\frac{1}{x_1^2}\right)\right)
\end{equation}
and
\begin{equation}
\beta= \frac{ix_1 +1}{ix_1-1}\left(1+\frac{1}{x_1^2}\right) \frac{\hat u {\cal C}}{2i} e^{-2ix_1}
\end{equation}
We are interested in the long time behaviour of the modes evaluated at $\eta_* \to 0$ implying that
\begin{equation}
v \approx \frac{\alpha +\beta}{x_*}
\end{equation}
We find that
\begin{equation}
v \approx \frac{\cal C}{x_*} \left(1+\hat u \frac{1+\frac{1}{x_1^2}}{i-\frac{1}{x_1}}\left(\cos x_1 - \frac{\sin x_1}{x_1}\right) e^{-ix_1}\right)
\end{equation}
Now defining $x_1= -\frac{k}{k_1}$ where $k_1=-\frac{1}{c_s \eta_1}$, we can study the limits $k\gg k_1$ and $k\ll k_1$. When $k$ is large, $\hat u$ goes to zero implying that
\begin{equation}
v (k\to \infty) = \frac{\cal C}{x_*}
\end{equation}
in an oscillatory manner. This correspond to a scale invariant spectrum $k^3 \vert v\vert ^2$. On the contrary we find that as $x_1\to 0$
\begin{equation}
v(k\to 0) = \frac{{\cal C}}{x_*}\left(1+\frac{\hat{u}x_1}{3}\right)
=\frac{{\cal C}}{x_*}\left(1-\frac{u}{3c_s k_1}\right)
\end{equation}
This implies that the power spectrum jumps from small to large $k$.

\subsection{Scalar-tensor features}
We are  interested in deriving analytical properties of  the power spectrum for $\cal R$ when $\rho_m\ll \rho_\phi$ and
$|W|\ll 1$. In this case we find that the source term $\Delta_\psi$ is regular and negligible. Moreover the speed of sound is not perturbed $\tilde c_s = c_s$.
The effect of the coupling function $A(\phi)$ appears at the level of its second derivative which is singular and behaves like a $\delta $ function. This leads to the following perturbation equation
\begin{equation}
v'' +\left(c_s^2 k^2 - \frac{z_A''}{z_A} \right) v=0
\end{equation}
with
\begin{equation}
z_A^2
=\frac{a^2}{H^2}\frac{1}{|1+W|}\tilde{z}^2
\end{equation}
where
\begin{equation}
\tilde{z}^2 \equiv
2X\left(\frac{\partial {\cal P}}{\partial X}+2X\frac{\partial^2 {\cal P}}{\partial X^2}\right)
=2X\left(\frac{\partial {\cal P}_{\rm{eff}}}{\partial X}+2X\frac{\partial^2 {\cal P}_{\rm{eff}}}{\partial X^2}\right)
\end{equation}
We have
\begin{eqnarray}
\nonumber
\frac{z_A''}{z_A}=\frac{a''}{a}+\frac{\tilde z''}{\tilde z}+2\frac{a'}{a}\frac{\tilde z'}{\tilde z}-{W'}\frac{\tilde z'}{\tilde z}-\frac{1}{2}{W''}-{W'}\frac{a'}{a}+\frac{3}{4}{W'^2}
\\
-2\frac{H'}{H}\frac{a'}{a}-2\frac{H'}{H}\frac{\tilde z'}{\tilde z}+\frac{H'}{H}W'-\frac{H''}{H}+2\frac{H'^2}{H^2}
\end{eqnarray}
%
%
where we find that
\begin{equation}
\label{Wsd}
W'' \supset \frac{\phi_1'}{\phi_1} \frac{k_c^2(\eta_1)}{k^2} \delta (\eta -\eta_1)
\end{equation}
where all the time dependent factors are evaluated at $\eta_1$. This allows one to identify
\begin{equation}
u_I(k)=\frac{\phi_1'}{\phi_1} \frac{k_c^2(\eta_1)}{2k^2}
\end{equation}
This is the first source of feature for scalar-tensor theories and it is due to the change of normalisation of the variable $z$ to the variable $z_A$. We notice that the coefficient of the Dirac function is scale-dependent. Another feature will also come from the coupling with matter through the matter coupling term  in the effective potential (\ref{app2}).

We are  interested in the terms containing $\tilde z$ and its derivatives :
\begin{equation}
\frac{z_A''}{z_A}=\frac{a''}{a}+\frac{1}{2}\frac{({\tilde z}^2)''}{{\tilde z}^2}-\frac{1}{2}{W''}+\mbox{  regular negligeable terms}
\end{equation}
Using
\begin{equation}
\frac{({\tilde z}^2)''}{{\tilde z}^2}=a{\cal H}\frac{1}{\tilde z^2}\frac{{\rm{d}}\tilde z^2}{{\rm{d}}t}+a^2\frac{1}{\tilde z^2}\frac{{\rm{d}}^2\tilde z^2}{{\rm{d}}t^2}
\end{equation}
and
\begin{eqnarray}
\nonumber
\frac{{\rm{d}}\tilde z^2}{{\rm{d}}t}=
-2\ddot\phi\dot\phi\left(\frac{\partial {\cal P}_{\rm{eff}}}{\partial X}+5X\frac{\partial^2 {\cal P}_{\rm{eff}}}{\partial X^2}+2X^2\frac{\partial^3 {\cal P}_{\rm{eff}}}{\partial X^3}\right)
\\
\label{eqqq}
+2X\dot\phi\left(\frac{\partial^2 {\cal P}_{\rm{eff}}}{\partial\phi\partial X}+2X\frac{\partial^3 {\cal P}_{\rm{eff}}}{\partial\phi\partial X^2}\right)
\end{eqnarray}
together with the Klein-Gordon equation (\ref{KGeff}), we can utilise 
\begin{equation}
\label{KGarrange}
\ddot\phi\left( \frac{\partial {\cal P}_{\rm{eff}}}{\partial X}+2X \frac{\partial^2 {\cal P}_{\rm{eff}}}{\partial X^2}\right) =
-\left(\frac{\partial {\cal P}_{\rm{eff}}}{\partial \phi} -2X\frac{\partial^2 {\cal P}_{\rm{eff}}}{\partial\phi\partial X}+3H\dot\phi\frac{\partial {\cal P}_{\rm{eff}}}{\partial X} \right)
\end{equation}
to replace $\ddot\phi$ in the previous equation (\ref{eqqq}). We find  that a term in $\frac{\partial {\cal P}_{\rm{eff}}}{\partial \phi}$ appears. If we derive a second time to compute $\frac{{\rm{d}}^2\tilde z^2}{{\rm{d}}t^2}$, we  obtain a long expression (see Appendix) which depends only on $X$, derivatives of ${\cal P}_{\rm{eff}}$ with respect to  $X$ up to the fourth order, mixed derivatives in $X$ and $\phi$, first derivatives of ${\cal P}_{\rm{eff}}$ with respect to $\phi$ and only one second-order derivative $\frac{\partial^2 {\cal P}_{\rm{eff}}}{\partial \phi^2}$. This  is the only term where $\frac{{\rm{d}}^2V_{\rm{eff}}}{{\rm{d}}\phi^2}$ is present, hence the only term where a singular second-derivative of the abruptly-evolving coupling with matter $\frac{{\rm{d}}^2 A}{{\rm{d}}\phi^2}$ appears. We find then that
\begin{equation}
\frac{{\rm{d}}^2\tilde z^2}{{\rm{d}}t^2}
\supset
-4X\frac{\partial^2 {\cal P}_{\rm{eff}}}{\partial \phi^2}
\frac{\frac{\partial {\cal P}_{\rm{eff}}}{\partial X}+5X\frac{\partial^2 {\cal P}_{\rm{eff}}}{\partial X^2}+2X^2\frac{\partial^3 {\cal P}_{\rm{eff}}}{\partial X^3}} {\frac{\partial {\cal P}_{\rm{eff}}}{\partial X}+2X \frac{\partial^2 {\cal P}_{\rm{eff}}}{\partial X^2}}
\end{equation}
so
\begin{eqnarray}
\frac{z_A''}{z_A}
\supset
a^2 \frac{{\rm{d}}^2V_{\rm{eff}}}{{\rm{d}}\phi^2}
\frac{\frac{\partial {\cal P}_{\rm{eff}}}{\partial X}+5X\frac{\partial^2 {\cal P}_{\rm{eff}}}{\partial X^2}+2X^2\frac{\partial^3 {\cal P}_{\rm{eff}}}{\partial X^3}}
{\left(\frac{\partial {\cal P}_{\rm{eff}}}{\partial X}+2X \frac{\partial^2 {\cal P}_{\rm{eff}}}{\partial X^2}\right)^2}
\end{eqnarray}
This formula can be used to evaluate the Dirac term in the perturbation equation for any  potential with discontinuous derivatives. If for example we have a canonical kinetic term ${\cal P}_{\rm{eff}}(\phi, X)=-X-V(\phi)$, we find that $u=a^2 \frac{{\rm{d}}^2V_{\rm{eff}}}{{\rm{d}}\phi^2}$ and it is straightforward to recover Starobinsky's result from section 4.1.

Hence we find that
\begin{equation}
\frac{{\rm{d}}^2V_{\rm{eff}}}{{\rm{d}}\phi^2}\supset\frac{\rho_m}{\phi_1}\delta(\phi-\phi_1)
\end{equation}
where
\begin{eqnarray}
u_{II}=-
\frac{\rho_m a_1 }{\phi_1\dot\phi_1}\delta(\eta-\eta_1)
\frac{\frac{\partial {\cal P}_{\rm{eff}}}{\partial X}+5X\frac{\partial^2 {\cal P}_{\rm{eff}}}{\partial X^2}+2X^2\frac{\partial^3 {\cal P}_{\rm{eff}}}{\partial X^3}}
{\left(\frac{\partial {\cal P}_{\rm{eff}}}{\partial X}+2X \frac{\partial^2 {\cal P}_{\rm{eff}}}{\partial X^2}\right)^2}
\end{eqnarray}
This is the second kind of feature which is present in scalar-tensor extensions of k-inflation coupled to matter. It is completely scale independent. The jump in the power spectrum due to this Dirac term will be dominant compared to the effect of the other Dirac term (\ref{Wsd}). \\
If we now consider a non trivial effective Lagrangian, such as the one inspired from the trapped brane case (\ref{app1}, \ref{app2})
\begin{equation}
{\cal P}_{\rm{eff}}(\phi, X)=\frac{1}{f(\phi)}\left(1-\sqrt{1+2Xf(\phi)}\right) -V_{\rm{eff}}(\phi)
\end{equation}
we obtain 
\begin{equation}
u_{II}=
\frac{\rho_m a_1}{\gamma_1\phi_1\dot\phi_1}\left(\frac{3}{2}-\frac{1}{2\gamma_1^2}\right)\delta(\eta-\eta_1)
\end{equation}
As a result, solving  the perturbation equation in a quasi de Sitter phase we find that the jump in the power spectrum is determined by
\begin{equation}
\label{saut}
\frac{v(k_c \ll k \ll k_1)}{v(k\to \infty)}\approx 1+ \frac{\eta_1 \phi'_1}{6\phi_1} \frac{k_c^2 (\eta_1)}{k^2}
+\frac{\rho_m a_1 \eta_1}{3\gamma_1\phi_1\dot\phi_1}\left(\frac{3}{2}-\frac{1}{2\gamma_1^2}\right)
\end{equation}
In particular we find that the spectrum jumps in the infrared. 

\subsection{Application}
We want to evaluate the impact of the feature for an effective model whose parameters are chosen to be the ones coinciding with that of the trapped brane case. 
As a result we have
\begin{equation}
-\frac{u_{II}}{3c_sk_1}\sim
-\frac{\rho_m a_1}{2\gamma_1\phi_1\dot\phi_1 k_1/\gamma_1}
\end{equation}
from  (\ref{e2}, \ref{e3}, \ref{e4}) which we can express  as
\begin{equation}
\label{est}
-\frac{u_{II}}{3c_sk_1}\sim
\frac{1}{2}\frac{1}{(2\pi)^3}\xi^{-1/2}g^2\frac{1}{\gamma_1}
\end{equation}
Recalling that \cite{Brax:2009hd, Silverstein:2003hf}
\begin{equation}
H \approx \frac{1}{\epsilon t}
\hspace{0.3cm}
\mbox{   and    }
\hspace{0.3cm}
\gamma \approx \frac{2m_{P}^{2}}{\lambda}\frac{1}{\epsilon}t^{2}
\end{equation}
where
\begin{equation}
f(\phi)=\frac{\lambda}{\phi^4}, 
\end{equation}
we deduce that
\begin{equation}
\gamma_1 \approx \frac{2m_P^2}{\lambda\epsilon^3 H_1^2}
\end{equation}
and we obtain that 
\begin{equation} 
\gamma_1 \sim \lambda\epsilon^3\frac{k_1^2/a_1^2}{2m_P^2}.
\end{equation}
Therefore we find
\begin{equation}
\label{use2}
-\frac{u_{II}}{3c_sk_1}\sim
\frac{1}{2}\frac{1}{(2\pi)^3}\epsilon^{-2}g^{5/2}\lambda^{-3/4}\frac{2m_P^2}{k_1^2/a_1^2}
\end{equation}
and typically for $k_1$ in the observable window, $\frac{2m_P^2}{k_1^2/a_1^2}$ varies from $10^{-6}$ to $10^5$. To satisfy the COBE normalisation, we can choose $\epsilon=10^{-1}$, $g=10^{-2}$ and $\lambda=10^9$. Therefore,
\begin{equation}
-\frac{u_{II}}{3c_sk_1}\sim
10^{-18}-10^{-7}
\end{equation}
This induces a jump which is  relatively small if $N=1$. For a choice of $N=10^6$, the background evolution is not affected (\ref{cond}) and  we can expect a noticeable effect in the power spectrum with an appropriate choice of $k_1$.
\begin{equation}
-\frac{u_{II}}{3c_sk_1}\sim
\frac{1}{2}\frac{1}{(2\pi)^3}N\epsilon^{-2}g^{5/2}\lambda^{-3/4}\frac{2m_P^2}{k_1^2/a_1^2}
\end{equation}
\begin{figure}[htbp]
\begin{center}
\label{figure1}
\includegraphics[width=7.5cm]{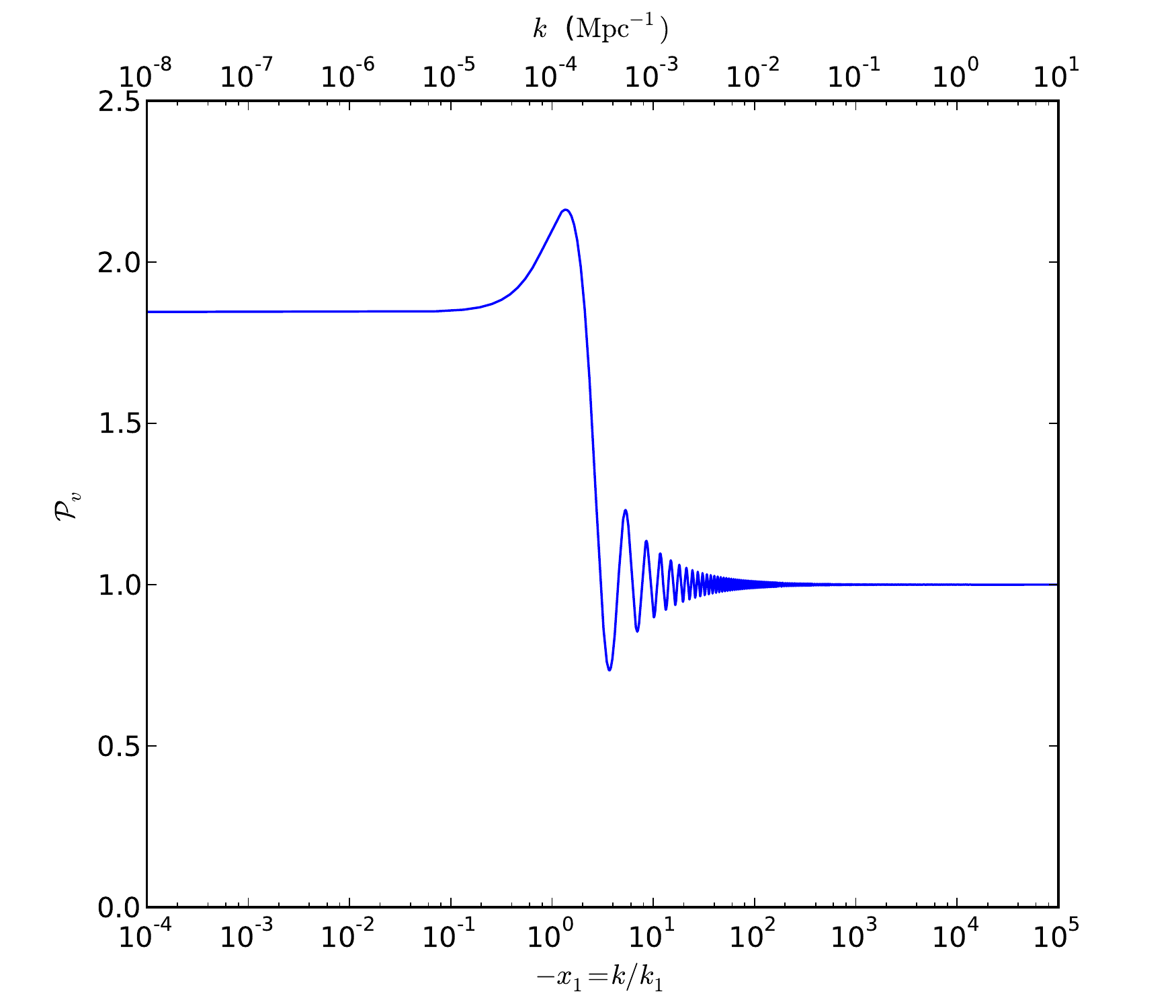}
\includegraphics[width=7.5cm]{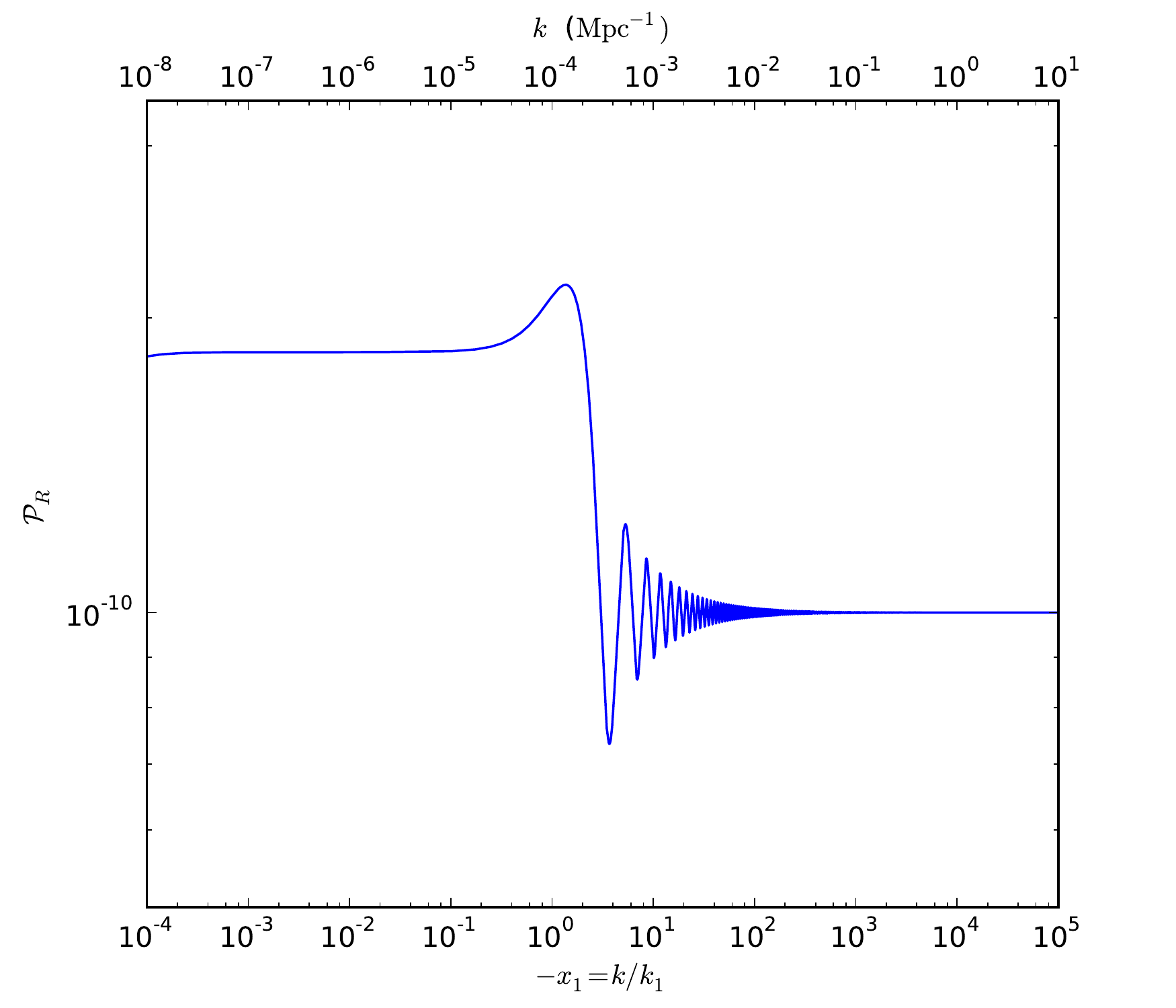}
\caption{Left : Power spectra for Mukhanov-Sasaki variable. Right : Power spectra for the curvature perturbation, for a feature in DBI inflation, with $\epsilon=10^{-1}$, $g=10^{-2}$ and $\lambda=10^9$ satisfying the COBE normalisation.}
\end{center}
\end{figure}
We have plotted the power spectrum of the Mukhanov-Sasaki variable and of the curvature perturbation $\cal R$ for $k_1=10^{-4} {\rm{Mpc}}^{-1}$, $\epsilon=10^{-1}$, $g=10^{-2}$, $\lambda=10^9$ and $N=10^6$. We have chosen $k_1$ in the observable window of the  Planck experiment \cite{planck} (roughly $10^{-4} {\rm{Mpc}}^{-1}-10^{-1} {\rm{Mpc}}^{-1}$). From (\ref{kc}), we can compute $k_c^2(\eta_1)/k_1^2$ :
\begin{equation}
\frac{k_c^2(\eta_1)}{k_1^2}=2^{-19/4}\pi^{-13/2}g^6\lambda^{-2}\epsilon^{-6}\zeta^{-7/2}\frac{\sqrt{2}m_P}{k_1/a_1}\frac{\phi_1}{\sqrt{2}m_P}
\approx 10^{-10} \ll 1
\end{equation}
implying that $k_c$ is tiny compared to $k_1$ and (\ref {saut}) is valid.

For such a small $k_c$, the curvature power spectrum is not much different from the power spectrum for $v$. 
For both spectra we observe a step in the spectrum which depends on the parameters of the model and  some additional oscillations.

\subsection{On other scales}
It is worth considering what happens for large scales such that $k\ll k_c$ i.e. $|W|\gg 1$ when  the energy density of matter is still negligible compared to the inflaton energy density $\rho_m\ll\rho_\phi$. In this case the perturbation equation reads
\begin{equation}
v_A'' + \left(k^2 \tilde c_s^2  -\frac{z_A''}{z_A}\right) v_A = 0
\end{equation}
with
\begin{equation}
\tilde {c_s^2}\approx c_s^2 \left(1- \frac{ W'}{{\cal H}W} \frac{2}{3(1+w_\phi)}\right)
\end{equation}
and
\begin{equation}
\frac{z_A''}{z_A}\approx \frac{9}{4}{{\cal H}}^2+\frac{3}{2}{\cal H'}-\frac{1}{2}\frac{A''}{A}+\frac{3}{4}\frac{A'^2}{A^2}-\frac{3}{2}{\cal H}\frac{A'}{A}
\end{equation}
The main contribution is $\frac{15}{4}{\cal H}^2$ much different from the usual $2{\cal H}^2$ in the de Sitter universe.  Moreover the speed of sound is greatly modified. This situation is too far from the de Sitter case to be of any phenomenological relevance.

\section{Conclusion}
In scalar-tensor extensions of k-inflation where matter is present, the perturbation equations are modified due to the coupling with matter. The energy density for matter can often be neglected compared to the energy density for the inflaton as the matter density is  diluted by the expansion of the universe.
On the other hand, when the coupling of matter changes abruptly along the inflationary trajectory, a jump in the power spectrum appears. The magnitude of this jumps depends crucially on the details of the k-inflationary models. Nevertheless for reasonable choices of parameters and in a scenario with a feature along a DBI inflationary trajectory, the change in the power spectrum could have observable consequences and would signal exotic physics in the inflationary scenario. The study of the phenomenology of this effect is left for future work.

\section{Acknowledgements}
We would like to thank C. van de Bruck, J. Martin, C. Ringeval, J. Weller for useful discussions and suggestions.

\appendix
\section{}

Here are some useful results related to section 4.3.

\begin{eqnarray}
\nonumber
\frac{z_A''}{z_A}=
\frac{a''}{a}+\frac{1}{2}a^2\frac{1}{\tilde z^2}\frac{{\rm{d}}^2\tilde z^2}{{\rm{d}}t^2}+\frac{1}{2}a{\cal H}\frac{1}{\tilde z^2}\frac{{\rm{d}}\tilde z^2}{{\rm{d}}t}-\frac{1}{4}a^2\left(\frac{1}{\tilde z^2}\frac{{\rm{d}}\tilde z^2}{{\rm{d}}t}\right)^2
+a\left({\cal H}-\frac{{\cal H'}}{{\cal H}}-\frac{W'}{2}\right)\frac{1}{\tilde z^2}\frac{{\rm{d}}\tilde z^2}{{\rm{d}}t}
\\
\label{grt}
-\frac{1}{2}{W''}+\frac{3}{4}{W'^2}+{W'}\left(\frac{{\cal H'}}{{\cal H}}-2{\cal H}\right)
+3{{\cal H}^2}-3{\cal H'}+2\left(\frac{{\cal H'}}{{\cal H}}\right)^2-\frac{{\cal H''}}{{\cal H}}
\end{eqnarray}
The last four terms sum up to zero in pure de Sitter case. We find that
\begin{eqnarray}
\nonumber
\frac{1}{\tilde z^2}\frac{{\rm{d}}\tilde z^2}{{\rm{d}}t}=\frac{\dot\phi}{X}
\frac{\left(\frac{\partial {\cal P}_{\rm{eff}}}{\partial \phi} -2X\frac{\partial^2 {\cal P}_{\rm{eff}}}{\partial\phi\partial X}+3H\dot\phi\frac{\partial {\cal P}_{\rm{eff}}}{\partial X} \right)
\left(\frac{\partial {\cal P}_{\rm{eff}}}{\partial X}+5X\frac{\partial^2 {\cal P}_{\rm{eff}}}{\partial X^2}+2X^2\frac{\partial^3 {\cal P}_{\rm{eff}}}{\partial X^3}\right)}{\left( \frac{\partial {\cal P}_{\rm{eff}}}{\partial X}+2X \frac{\partial^2 {\cal P}_{\rm{eff}}}{\partial X^2}\right)^2}
\\
+\dot\phi \frac{\frac{\partial^2 {\cal P}_{\rm{eff}}}{\partial\phi\partial X}+2X \frac{\partial^3 {\cal P}_{\rm{eff}}}{\partial\phi\partial X^2}}
{\frac{\partial {\cal P}_{\rm{eff}}}{\partial X}+2X \frac{\partial^2 {\cal P}_{\rm{eff}}}{\partial X^2}}
\end{eqnarray}
For simplicity and conciseness, we define :
\begin{eqnarray}
\nonumber
M=\left(\frac{\partial {\cal P}_{\rm{eff}}}{\partial \phi} -2X\frac{\partial^2 {\cal P}_{\rm{eff}}}{\partial\phi\partial X}+3H\dot\phi\frac{\partial {\cal P}_{\rm{eff}}}{\partial X} \right)
=\tilde{M}+3H\dot\phi\frac{\partial {\cal P}_{\rm{eff}}}{\partial X} 
\\
\nonumber
N=\left(\frac{\partial {\cal P}_{\rm{eff}}}{\partial X}+5X\frac{\partial^2 {\cal P}_{\rm{eff}}}{\partial X^2}+2X^2\frac{\partial^3 {\cal P}_{\rm{eff}}}{\partial X^3}\right)
\\
\nonumber
Q=\left( \frac{\partial {\cal P}_{\rm{eff}}}{\partial X}+2X \frac{\partial^2 {\cal P}_{\rm{eff}}}{\partial X^2}\right)
\end{eqnarray}
and we express
\newpage
\begin{eqnarray}
\nonumber
\frac{1}{\tilde z^2}\frac{{\rm{d}}^2\tilde z^2}{{\rm{d}}t^2}=
3\ddot\phi \left( \frac{\partial^2 {\cal P}_{\rm{eff}}}{\partial\phi\partial X}
+ \frac{\partial^3 {\cal P}_{\rm{eff}}}{\partial\phi\partial X^2}\right)\frac{1}{Q}
+\dot\phi \dot{X}\left( 3\frac{\partial^3 {\cal P}_{\rm{eff}}}{\partial\phi\partial X^2}+2X \frac{\partial^4 {\cal P}_{\rm{eff}}}{\partial\phi\partial X^3}\right)\frac{1}{Q}
\\
\nonumber
-2X\left( \frac{\partial^3 {\cal P}_{\rm{eff}}}{\partial\phi^2\partial X}+2X \frac{\partial^4 {\cal P}_{\rm{eff}}}{\partial\phi^2\partial X^2}\right)\frac{1}{Q}
-2 \left(\frac{\partial_\phi \tilde{M}}{M}+\frac{\partial_\phi N}{N}-\frac{\partial_\phi Q}{Q}\right)\frac{M.N}{Q^2}
\\
\nonumber
+\frac{\dot\phi}{X} \left(3\dot{H}\dot\phi \frac{\partial {\cal P}_{\rm{eff}}}{\partial X} +3H\ddot\phi \frac{\partial {\cal P}_{\rm{eff}}}{\partial X} +3H\dot\phi^2 \frac{\partial^² {\cal P}_{\rm{eff}}}{\partial\phi \partial X} + 3H\dot\phi \dot{X} \frac{\partial^2 {\cal P}_{\rm{eff}}}{\partial X^2}\right)  \frac{N}{Q^2}
\\
\label{expr}
+2\ddot\phi \left(\frac{\partial_X \tilde{M}}{M}+\frac{\partial_X N}{N}-\frac{\partial_X Q}{Q}\right)\frac{M.N}{Q^2}
+\frac{\ddot\phi}{X} \frac{M.N}{Q^2}
\end{eqnarray}
with
\begin{eqnarray}
\nonumber
\partial_\phi\tilde{M}=\frac{\partial^2 {\cal P}_{\rm{eff}}}{\partial \phi^2} -2X\frac{\partial^3 {\cal P}_{\rm{eff}}}{\partial\phi^2\partial X}
\end{eqnarray}
\begin{eqnarray}
\nonumber
\partial_X\tilde{M}=-\frac{\partial^2 {\cal P}_{\rm{eff}}}{\partial \phi\partial X} -2X\frac{\partial^3 {\cal P}_{\rm{eff}}}{\partial\phi\partial X^2}
\end{eqnarray}
\begin{eqnarray}
\nonumber
\partial_\phi {N}=
\frac{\partial^2 {\cal P}_{\rm{eff}}}{\partial\phi\partial X}+5X\frac{\partial^3 {\cal P}_{\rm{eff}}}{\partial\phi\partial X^2}+2X^2\frac{\partial^4 {\cal P}_{\rm{eff}}}{\partial\phi\partial X^3}
\end{eqnarray}
\begin{eqnarray}
\nonumber
\partial_X {N}=
6\frac{\partial^2 {\cal P}_{\rm{eff}}}{\partial X^2}+9X\frac{\partial^3 {\cal P}_{\rm{eff}}}{\partial X^3}+2X^2\frac{\partial^4 {\cal P}_{\rm{eff}}}{\partial X^4}
\end{eqnarray}
\begin{eqnarray}
\nonumber
\partial_\phi {Q}=\frac{\partial^2 {\cal P}_{\rm{eff}}}{\partial\phi\partial X}+2X \frac{\partial^3 {\cal P}_{\rm{eff}}}{\partial\phi\partial X^2}
\end{eqnarray}
\begin{eqnarray}
\nonumber
\partial_X {Q}=\frac{3\partial^2 {\cal P}_{\rm{eff}}}{\partial X^2}+2X \frac{\partial^3 {\cal P}_{\rm{eff}}}{\partial X^3}
\end{eqnarray}
In equation (\ref{expr}), we can replace again $\ddot\phi$ using the Klein-Gordon equation (\ref{KGarrange}), we can use the Friedmann equation (\ref{Feq}) to replace $H$ and we can use the Hamilton-Jacobi equation (\ref{KGmeq}) to replace $\dot{H}$ where
\begin{equation}
\label{Feq}
H^2
=\frac{8\pi G_N}{3}\left(-{\cal P_{\rm{eff}}} +2X \frac{\partial {\cal P}_{\rm{eff}}}{\partial X}  \right)
\end{equation}
The Hamilton-Jacobi equation is obtained by combining the Klein-Gordon equation and the derivative of the Friedmann equation
\begin{equation}
\label{KGmeq}
\dot\phi \frac{\partial {\cal P}_{\rm{eff}}}{\partial X}=\frac{1}{4\pi G_N}\frac{{\rm{d}}H}{{\rm{d}}\phi}
\end{equation}
If we use this equation to derive ${{\cal H''}}$, we can check that no $\frac{\partial^2{\cal P}_{\rm{eff}}}{\partial \phi^2}$ term appears, so that the term $-{\cal H''}/{\cal H}$ in (\ref{grt}) contains no singularity.

\end{document}